# High-pressure melting and elastic behavior of vanadium and niobium based on *ab initio* and machine learning molecular dynamics

Hao Wang[1], Dan Wang[1, 3], Long Hao[1], Jun Li[1], Hua Y. Geng[1, 2*]

[1] *National Key Laboratory of Shock Wave and Detonation Physics, Institute of Fluid Physics, CAEP, Mianyang 621900, People's Republic of China*

[2] *HEDPS, Center for Applied Physics and Technology, and College of Engineering, Peking University, Beijing 100871, People's Republic of China*

[3] *Institute of Atomic and Molecular Physics, College of Physics, Sichuan University, Chengdu 610065, People's Republic of China*

**Abstract:** Under high pressure, the group-VB transition metals vanadium (V) and niobium (Nb) exhibit simple crystal structures but complex physical behaviors, such as anomalous compression-induced softening and heating-induced hardening (CISHIH). Meanwhile, the impact of lattice thermal expansion-induced softening at elevated temperatures on HIH is yet to be investigated. Therefore, this study utilized *ab initio* (AIMD) and machine learning molecular dynamics (MLMD) to investigate the melting and abnormal mechanical softening–hardening behaviors of V and Nb under high pressure. Simulations reveal that the high-temperature Pnma phase of Nb reported in previous experimental studies is highly susceptible to mechanical instability and reverts to the body-centered cubic (BCC) phase. This discovery prompted a revised determination of the high-pressure melting line of Nb. The melting temperature of Nb significantly exceeds the existing theoretical and experimental estimate compared with that of V. AIMD simulations demonstrate that atomic thermal displacements have a greater influence on the HIH of V and Nb than pure electron temperature effects. In addition, the temperature-dependent anomalous elastic properties of V and Nb were investigated within a pressure range of 0–250 GPa using MLMD. The mechanical properties of V and Nb transitioned from HIH to heating-induced softening, elucidating the competition between thermal-expansion-induced softening and HIH. This study advances fundamental understanding of V and Nb physics, providing crucial theoretical foundations for establishing accurate equations of state and constitutive models for these metals.

**Keywords:** melting; anomalous elastic behavior; vanadium and niobium; high pressure; machine learning potential

---

\* *Corresponding author. E-mail:* s102genghy@caep.cn

## 1. Introduction

Transition metals have broad applications in industrial and aerospace fields, and researchers have investigated their physical properties under high pressures. Although their crystal structures are considerably simple owing to unfilled $d$-orbital electrons, these metals often possess unusual physical properties under high pressure, such as Fermi surface nesting, electronic topological transitions, and Jahn–Teller effects [1, 2]. Among the elements V and Nb are the most typical representations.

Early studies have suggested significant differences in the shock and static melting behaviors of V [3, 4]. However, recent improvements in experimental measurement accuracy have reestablished and unified the shock and static melting curves of V, contradicting previously reported discrepancies between the two [5, 6]. Notwithstanding the extensive research on the melting of V, few studies have been reported on the high-pressure melting of Nb, which belongs to the same group of high-melting metals, particularly regarding shock melting. In 2020, Errandonea [7] conducted melting experiments and theoretical studies on Nb under static high pressure for the first time and discovered a transition from the BCC to the Pnma orthogonal phase at high temperatures.

The mechanical behaviors of V and Nb under high pressure have attracted significant attention owing to their anomalous compression-induced softening and heating-induced hardening (CISHIH) properties [8-13]. Thus, establishing new mechanical constitutive models under high pressure is crucial[14]. The earliest indications of elastic anomalies in V and Nb were observed in the abnormal softening behavior of the transverse acoustic branch of the phonon spectrum under high pressure [15, 16]. Landa $et\ al.$[8], using first-principle calculations, discovered that the shear elastic constant $C_{44}$ of V and Nb exhibited an abnormal CIS under high pressure, which they believed was due to changes in the electronic structure. Koči [11] and Liu $et\ al.$ [17] suggest that the anomalies may be related to Fermi surface nesting. Wang $et\ al.$ [12, 13] discovered that V and Nb exhibited abnormal HIH. The CISHIH of V was confirmed in shock-sound velocity experiments and density functional theory (DFT) calculations [18]. However, the impact of lattice thermal expansion-induced softening at elevated temperatures on HIH behavior has not been investigated.

Molecular dynamics simulation is a useful method for studying material properties under extreme conditions. Existing classical potentials [18-22] struggle to accurately describe the physical properties of transition metals such as V and Nb, which exhibit localized electronic behavior under high pressures. Substantial computational resources are required for ab initio molecular dynamics (AIMD) simulations to achieve accurate results. Recently, the development of machine learning has combined the advantages of classical potential and AIMD. Machine learning potentials, such as the Gaussian approximation potential (GAP) [23, 24], moment tensor potential (MTP) [25, 26], and deep learning potential (DeepMD) [27, 28], have been extensively utilized in material property simulations [29-31]. Byggmästar *et al.*[32] constructed GAP models for V and Nb, and investigated their melting behavior under high pressure. Srinivasan *et al.*[33] and Zotov *et al.*[34] explored methods for developing more reliable MTP models to describe the mechanical and thermal properties of Nb. However, the applicability of these MTP potentials is currently limited to studies at ambient pressure, precluding their use in high-pressure scenarios.

In this study, we trained the MTP of V and Nb using a dataset generated from VASP's on-the-fly force field training. The stability of the high-temperature Pnma phase of Nb was also investigated. We then used the MTP to predict the melting curves of V and Nb under high pressure and compared them with existing experimental and theoretical results. Furthermore, we predicted changes in the elastic properties and sound velocity of V and Nb with temperature. We also analyzed the competition between HIH and heat-induced softening (HIS) and determined the HIH–HIS boundary.

## 2. Methods

### 2.1. Training strategy

MTP training was performed using the software named *MLIP*-v2 [26]. The normal machine learning potential generation is divided into two steps: (1) establishing complete training and test datasets through DFT calculations, and (2) training and generating potential functions using machine learning methods. However, this solution may require extensive DFT calculations. Therefore, a new training method called active learning, which enables the on-the-fly generation and accumulation of training data, was developed. *MLIP*-v2 software has been used to develop a corresponding active learning flowchart and several applications [26, 35-37]. However, for the pressure and temperature ranges considered herein, the MTP's active training strategy is inefficient.

Figure 1 shows the on-the-fly machine-learning force field generation framework implemented in *VASP*-6 [38, 39], which was used to generate the training set required for the MTP. The VASP training framework can run directly on graphics processing unit hardware, making it more efficient than other CPU-based training frameworks. Within targeted pressure regimes, datasets exhibiting sufficient structural diversity to encompass relevant configurational space may serve as standardized training sets for interatomic potentials [29]. The primary rationale for the abandonment of VASP's machine-learning force field (VASP-MLFF) was its excessive memory consumption during multi-pressure dataset retraining. This inefficiency is likely attributable to inherent limitations in the descriptors and algorithms of our VASP version. Consequently, we have adopted the MTP potential, which is more computationally efficient.

For the on-the-fly generation, a supercell containing 128 atoms with a time step of 1 fs was used. Isothermal–isobaric ensemble (NPT) was used to obtain structures with different strains at different temperatures. The training pressure range was 0–250 GPa, with intervals of 50 GPa, and each pressure point was independent in the on-the-fly training. The training temperature ranged from 100 K and continued until the structure completely turned into a liquid. The heating rate was 0.05 K/fs. The training set for V (Nb) comprised 5700 (6900) configurations. For the MTP training, the cut-off radius for V (Nb) was set to 6.5 (7.5) Å, with energy, force, and stress weights of 1, 0.01, and 0.1, respectively. The level of MTP was selected *lev*-22[26].

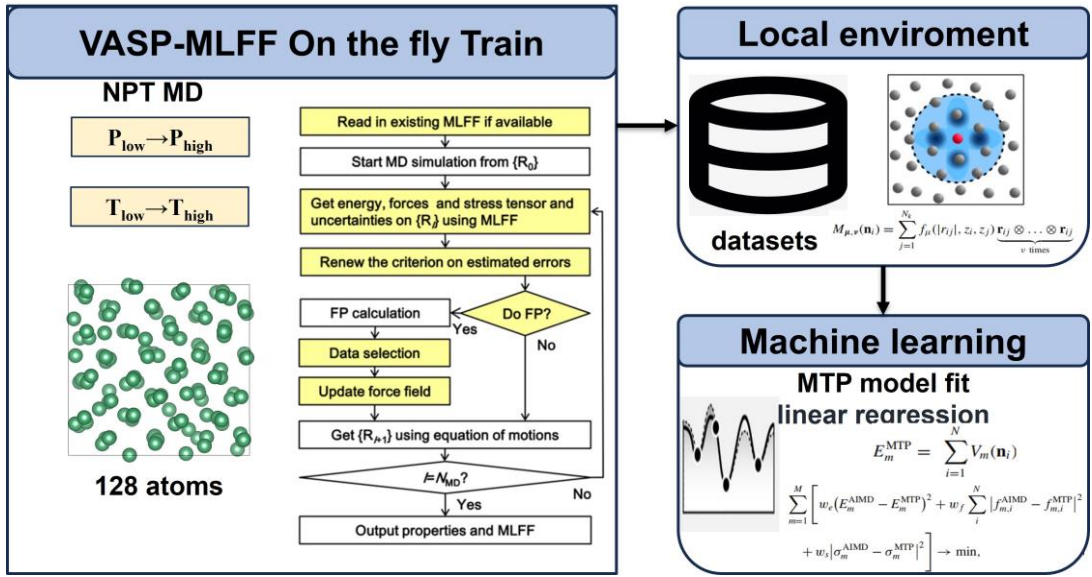

**Fig. 1**. (Color online) Flow chart of MTP training for V and Nb. On-the-fly training dataset generation scheme was obtained from Ref. [38]. Theoretical details of MTP can be found in the references [25, 26]. VASP-MLFF represents the VASP machine learning force field.

## 2.2. *Ab initio* calculations

In this study, DFT calculations were performed using the *VASP* package (version 6.3) [40, 41], and a plane-wave basis set was employed with a kinetic energy cut-off of 600 eV for V and Nb. The electron–core interaction was described by a projector-augmented wave pseudopotential [42], comprising 13 valence electrons for V and Nb. The electron exchange correlation was handled using a generalized gradient approximation with Perdew–Berke–Ernzerhof parameterization [43].

AIMD results were used to verify the accuracy of the MTP. The supercell size was 4×4×4 (128 atoms), and the K-point sampling grid in the irreducible Brillouin zone was 3×3×3. In the case to validate the stability of the Pnma structure of Nb, a large $4 \times 4 \times 7$ supercell (448 atoms) and a single $\Gamma$ point were employed. The energy convergence criterion was set to $10^{-5}$ eV/cell, and the time step was 1 fs. The NPT ensemble was used to obtain the lattice constant at a given temperature and pressure, and the total time was 3 ps. Temperature-dependent effective potentials [44, 45] and the stress-strain method [46] were applied using the *TDEP* and *MyElas* packages [47] to calculate the finite-temperature lattice dynamics and elastic constants. A canonical ensemble (NVT) was used to ensure the invariance of the volume and temperature in the AIMD calculations, with a total time of 20 ps.

The melting curves of bcc V and Nb were calculated using the Z-method (ZM) implemented in the AIMD simulations. A 6×6×6 (432 atoms) supercell with a single $\Gamma$ point was used. Two melting points of V and Nb were calculated for different densities, respectively. A microcanonical ensemble (NVE) was used to equilibrate the system.

To characterize the structure after molecular dynamics (MD) simulation, we averaged the lattice constants and atomic positions from the last 1000 steps of the MD to obtain the equilibrium structure. Additionally, we calculated its X-ray diffraction pattern (with a X-ray wavelength λ=0.3738 ) using the VESTA program[48] and analyzed the crystalline phase by integrating the polyhedral method via OVITO[49].

## 2.3. Machine learning molecular dynamics simulations

MLMD simulations with MTP potentials were performed using the *LAMMPS* package [50]. Melting calculations were conducted using NVT ensemble with two-phase method and a 60×20×20 supercell (48000 atoms). Supercell of 10×10×10 containing 2000 atoms was used for other calculations. The step size of the MD simulation was 1 fs, and the total time was sufficiently long to ensure that the thermodynamic quantities were in equilibrium. Full relaxation in the NPT ensemble was used to calculate the lattice constant, and the pressure (temperature) damping parameters for the thermostat were set to 1.0 (0.1). An NVT ensemble was then used to obtain stress when the elastic constant was calculated. The strain values used for the elastic constant calculation of V and Nb were -0.02, -0.01, 0, 0.01, and 0.02.

## 3. Results and discussion

### 3.1. Benchmarks of the moment tensor potential

The root mean square errors of energy, force, and stress for the V (Nb) training set were 3.2 (3.5) meV/atom, 0.21 (0.17) eV/Å, and 0.3 (0.2) GPa, respectively. Figure 2 shows a comparison of the energies, forces, and stresses between the DFT and MTP for all test data. The root mean square errors of the energy, force, and stress for the V (Nb) testing set were 9 (7) meV/atom, 0.23 (0.18) eV/Å, and 0.32 (0.23) GPa, respectively. Figure 3 verifies the reliability of the MTP in describing the physical characteristics of V and Nb through the equation of state, elastic properties, and lattice dynamics results. The MTP was consistent with AIMD and the experimental results in terms of the P–V curves [51-54] and lattice dynamic properties [15, 55]. We observed that the predictions of niobium's phonon spectra by the MTP and AIMD show discrepancies at the edge of the Brillouin zone, which may affect phase stability predictions. We evaluated the impact of these differences on our results by comparing the free energies calculated using both methods. The free energy difference between DFT (-10.252 eV/atom) and MTP (-10.249 eV/atom) was particularly small, which generally will not significantly affect our computational conclusions.

Regarding the elastic properties, the MTP shows excellent agreement with DFT and other MLIPs at 0 GPa (Table 1 and Fig. 3(b) and (e)). In addition, it inherited the deviation between the DFT and the experiment [56-58] in single-crystal elastic constants. The detailed comparison of the elastic properties with experimental values [57, 59] provided in Fig. 4. The MTP successfully captures the qualitative trends in elastic behavior. Notably, in contrast to earlier finite-temperature computational studies [12, 13] for V and Nb, the MTP accurately reproduces the softening behavior of the shear elastic constant $C'$. This finding underscores the efficacy of the MTP in providing a reliable framework for modeling the temperature-dependent elastic properties of V and Nb. For comparison, the classical potential of V [19, 20] cannot accurately describe its elastic properties at finite temperatures. Although the MEAM potential of Nb [21, 22] performs effectively at zero pressure, it is deemed unsuitable for high-pressure scenarios.

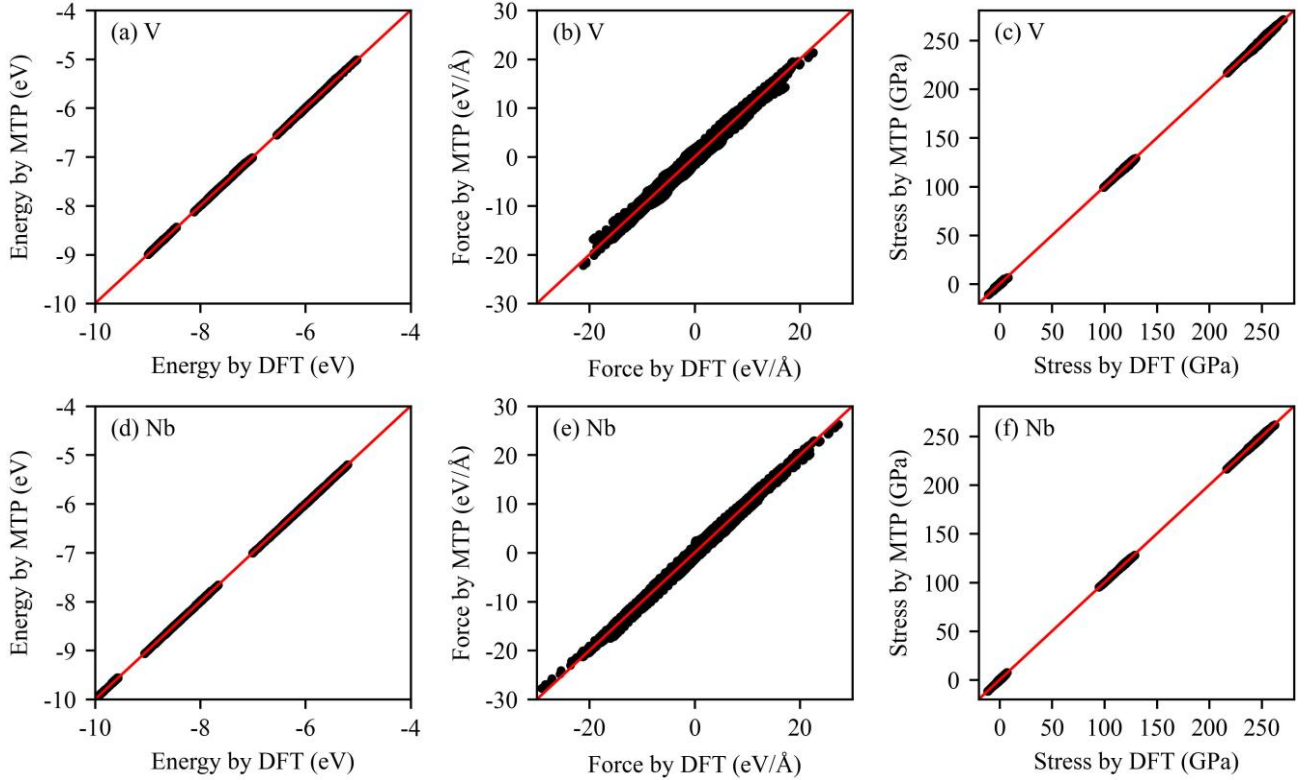

**Fig. 2**. (Color online) Comparisons of energies, forces, and stresses between DFT and MTP for all test data at different pressures and temperatures.

While GAP [32] and MTP exhibit comparable accuracy, MTP (13250 atoms·step/s) offers a computational efficiency boost of approximately 9 times that of GAP (1450 atoms·step/s). Testing was conducted using two Intel Xeon Silver 4114 processors with 20 cores and 40 threads (10 cores and 20 threads per processor). In addition, at elevated pressures, the prediction of elastic properties from GAP exhibited a notable deviation from that obtained from AIMD. For example, at 150 GPa and 300 K, the elastic constants ($C_{11}$, $C_{12}$, and $C_{44}$) of Nb obtained from AIMD (MTP) were 922 (928), 502 (516), and 144 (138) GPa, respectively. The GAP values were 1078, 518, and 207 GPa, respectively.

**Table 1**. Basic static properties (0 K) of BCC V and Nb obtained by our MTP and DFT calculations

and previous interatomic potential and experimental measurements (4.2 K). The unit of the lattice constant ($a_0$) is Å, and that of the elastic constants ($C_{ij}$) is GPa.

|   |   | MTP | DFT | MTP[60] |   | GAP[32] | MEAM[19] | EAM[20] | Expt.[61] |
|---|---|-----|-----|---------|---|---------|----------|---------|-----------|
| V | $a_0$ | 2.998 | 2.998 | 3.00 |   | 2.997 | 3.03 | 3.03 | 3.024 |
|   | $C_{11}$ | 278 | 270 | 261 |   | 271 | 232 | 232 | 237 |
|   | $C_{12}$ | 149 | 139 | 144 |   | 145 | 119 | 119 | 121 |
|   | $C_{44}$ | 23 | 24 | 23 |   | 23 | 46 | 46 | 47 |

|   |   | MTP | DFT | MTP[34] | MTP[62] | GAP[32] | MEAM[22] | EAM[21] | Expt.[63] |
|---|---|-----|-----|---------|---------|---------|----------|---------|-----------|
| Nb | $a_0$ | 3.307 | 3.308 | 3.322 | 3.323 | 3.308 | 3.322 | 3.308 | 3.303 |
|   | $C_{11}$ | 264 | 247 | 247 | 269 | 243 | 250 | 244 | 253 |
|   | $C_{12}$ | 135 | 133 | 139 | 127 | 137 | 135 | 136 | 134 |
|   | $C_{44}$ | 14 | 19 | 18 | 16 | 13 | 21 | 32 | 30.9 |

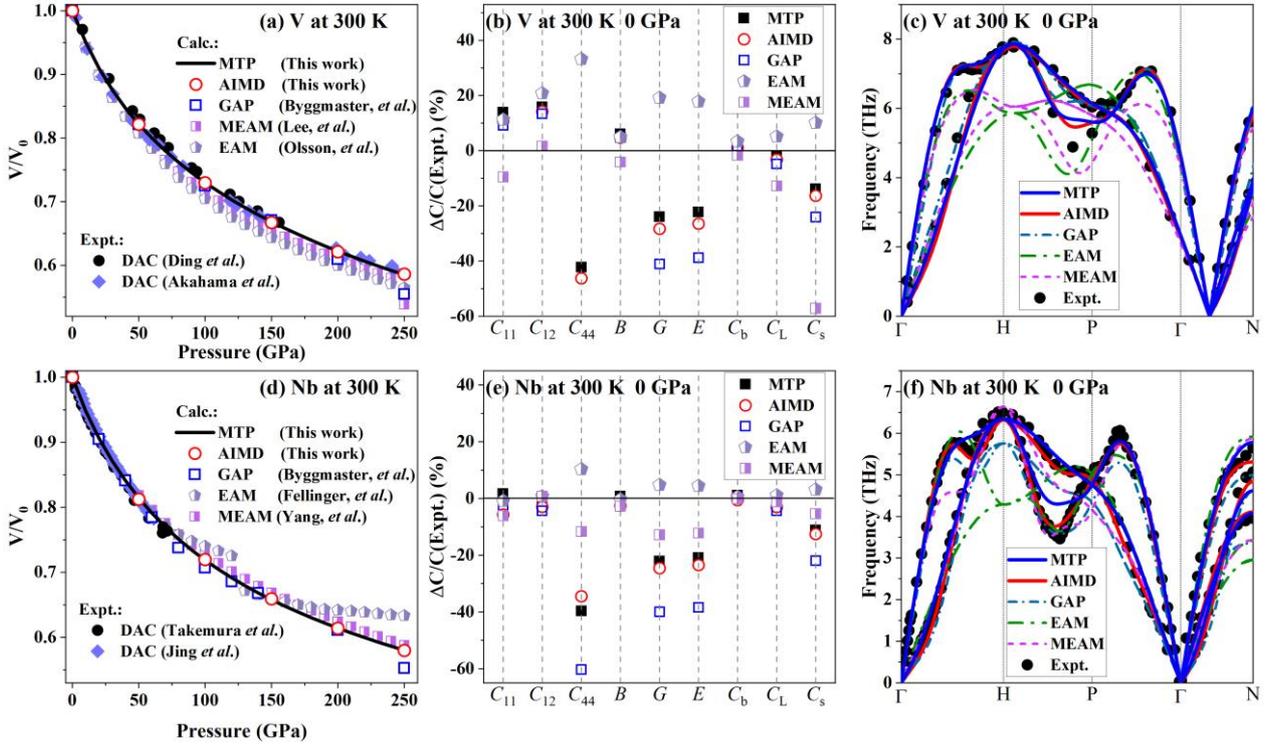

**Fig. 3**. (Color online) P–V curves, elastic properties, and phonon spectra of V and Nb at 300 K. MTP results are compared with those of AIMD, other interatomic potentials [19-22, 32], and experimental data [15, 51-57].

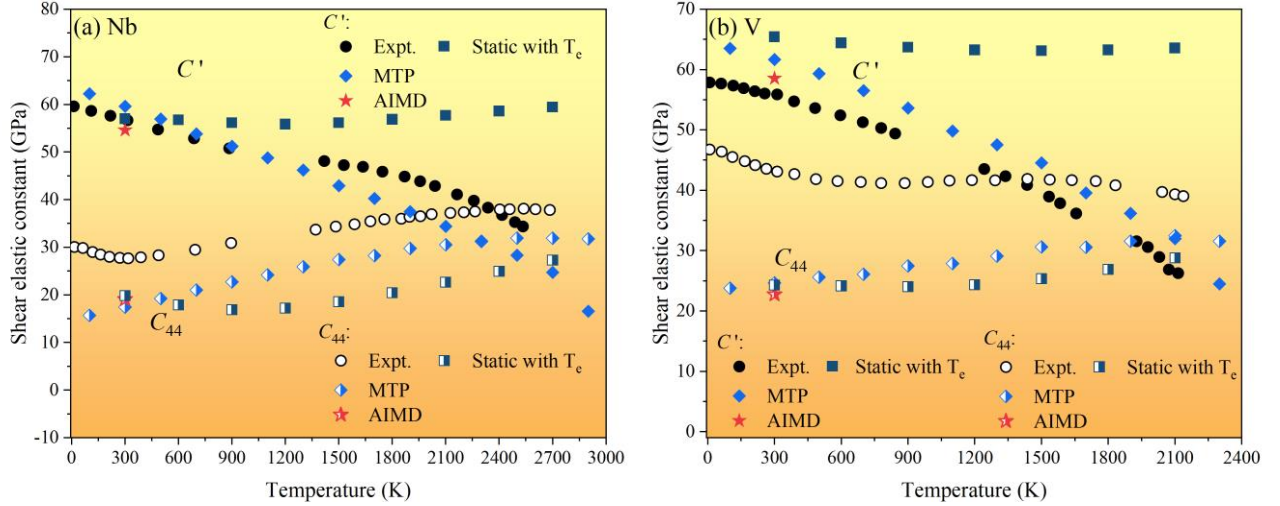

**Fig. 4**. (Color online) Calculated variation of elastic constants of bcc Nb and V as a function of temperature at 0 GPa. (a) $C_{44}$ and $C'$ of Nb; (a) $C_{44}$ and $C'$ of V. Solid and open circles denote the experimental $C'$ and $C_{44}$ values[57, 59], respectively. Other Solid and half-filled symbols represent the theoretical $C'$ and $C_{44}$ results, respectively. SCF (with electron temperature $T_e$) data references literature [12] and [13].

## 3.2. The Stability of the Pnma Phase in Nb

Before investigating the temperature dependence of the elastic properties in V and Nb, it is essential to accurately define the solid-phase regions in their high-temperature and high-pressure phase diagrams. Recent experimental studies have demonstrated that the high-pressure rhombohedral (RH) phase previously reported in V may arise from weak lattice distortions of the BCC structure [54, 64]. Therefore, within the pressure range investigated in this study (0–250 GPa), V retains the BCC phase as the stable solid-state structure. However, Errandonea [7] highlighted that Nb undergoes a BCC → Pnma → liquid transition during heating at high pressure. This phase transition mechanism is particularly rare, as most metals with high-temperature phase transitions tend to adopt higher-symmetry structures.

Therefore, in this study, we investigated the high-temperature stability of the Pnma phase. We relaxed the ideal Pnma structure in its stable region using AIMD. The final equilibrium structure was determined by considering the structure of the last thousand steps and statistically averaging the lattice constants and coordinates. The theoretical X-ray diffraction (XRD) pattern of the equilibrium structure was then calculated, as shown in Fig. 5. The 50 GPa Pnma structure is based on the Errandonea experimental value [7]. A comparison of the theoretical XRD patterns confirmed that the Pnma structure transformed into a BCC structure, particularly in the NPT simulation. In addition, we discovered that the solid structure before melting had already transformed into another structure when the Z method was used to calculate the melting point of the Pnma structure (see the Supplementary Material [65]). This suggests that the Pnma phase may not represent the high-temperature-stable phase structure.

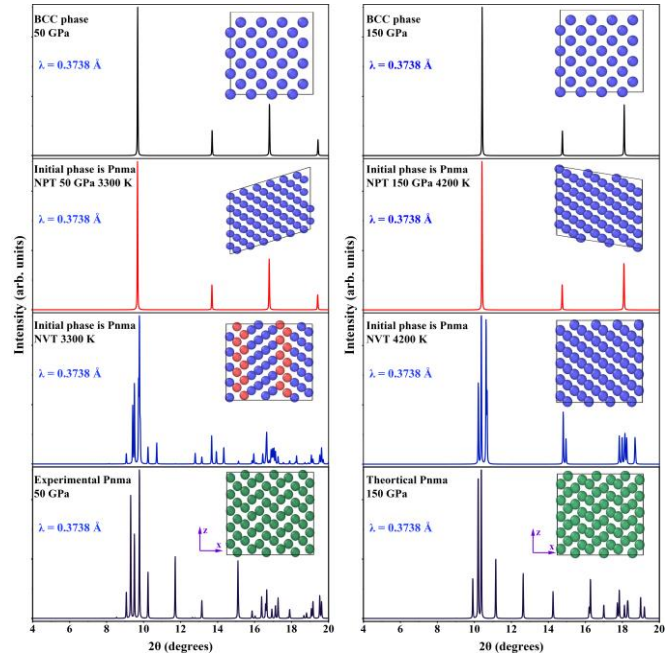

**Fig. 5**. (Color online) Theoretical XRD patterns of the optimized structures of the Pnma phase at 50 and 150 GPa, obtained via NPT and NVT AIMD simulations. The standard theoretical XRD patterns of the Pnma and BCC phases are also presented in the figure. The initial structure of the Pnma phase at 50 GPa is constructed based on experimental data reported by Errandonea [7]. λ is the X-ray wavelength. The inset figure is the optimized final structure. Based on tetrahedron method analysis, the blue atoms represent the BCC structure, green represents the Pnma structure, and red may indicate the interface between two different crystallographic orientations of the BCC structures in MD simulation.

Furthermore, we performed constrained NPT ensemble simulations with fixed lattice geometry (unconstrained NPT caused complete reversion to BCC). Subsequent NVT simulations of these equilibrated structures revealed pronounced dynamical instabilities in the Pnma phase at high temperature, as confirmed by imaginary phonon frequencies along the Γ–Z and Γ–Y paths (Fig. 6(a)). Conversely, consistent with Errandonea's methodology in which NVT ensemble simulations were performed directly on statically optimized structures, our phonon dispersion calculations demonstrate the dynamical stability of such a prepared Pnma phase at elevated temperatures (Fig. 6(b)). Importantly, these NVT simulations produced significant stress anisotropy ($\Delta\sigma_{max}$ = 10-20 GPa along the principal directions), deviating markedly from hydrostatic conditions. These findings collectively suggest that the Pnma phase, as reported in the DAC experiment [7], may achieve metastability under non-hydrostatic conditions.

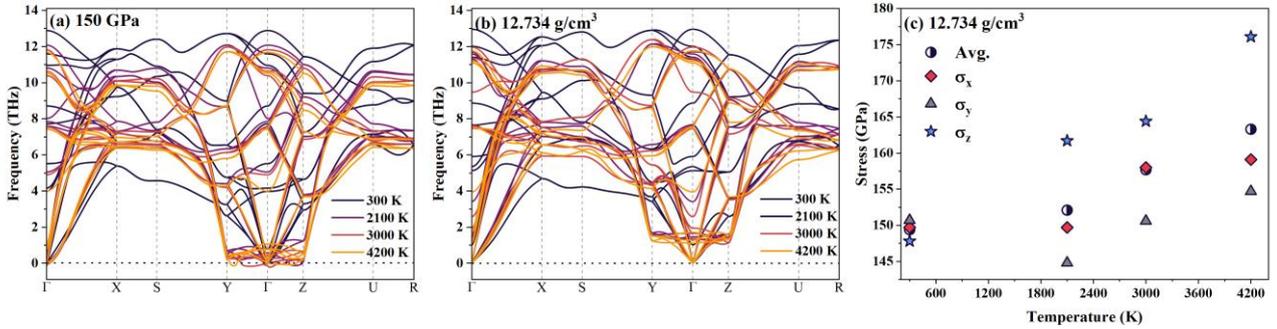

**Fig. 6**. (Color online) Calculated phonon dispersion relations of the Pnma phase under different conditions: (a) at 150 GPa and (b) at a density of 12.7 g/cm³ with a non-hydrostatic condition. (c) Stress anisotropy exhibited by the Pnma phase at fixed volume across temperatures, as shown in (b). Half-filled circles denote the mean stress from NVT simulations, whereas filled diamonds, triangles, and pentagons represent the principal stresses along the x-, y-, and z-directions, respectively.

### 3.3. Melting Curves

Following the clarification of the high-temperature instability of the Pnma-phase in Nb, we calculated the melting curves of V and Nb using the two-phase method based on MTP (with NVT ensemble) and the Z method based on AIMD (with NVE ensemble). The results are shown in Fig. 7. The MTP results agreed well with the AIMD results. The melting curves of V and Nb can be fitted using the Simon–Glatzel equation [66] $T_m = T_{m0}(P/a+1)^{1/b}$, where $T_{m0}$ is the melting point at zero pressure, and $a$ and $b$ are the fitting parameters. The coefficients $a$ and $b$ for V (Nb) were 19.06 (23.31 GPa⁻¹) and 2.82 (2.41). The melting curve of V given by the MTP within 200 GPa was in excellent agreement with the experimental values [5, 6], as shown in Fig. 7(a). The $T_{m0}$ and initial Clapeyron slope given by MTP were 2096 K and 39 K/GPa, respectively. They were consistent with the previous experimental values of 2183 K and 31.4–33.9 K/GPa [5, 6, 67]. Our MTP results predict that V begins to melt at 207 GPa under shock compression and is completely melted at 257 GPa. This result is consistent with Dai's shock–sound velocity data [4]. The melting curves calculated in this study and by the AIMDs in other studies [5, 68] also validated the reliability of our MTP in predicting the melting properties of V. By contrast, the melting temperature was significantly underestimated by the reported GAP for V [32].

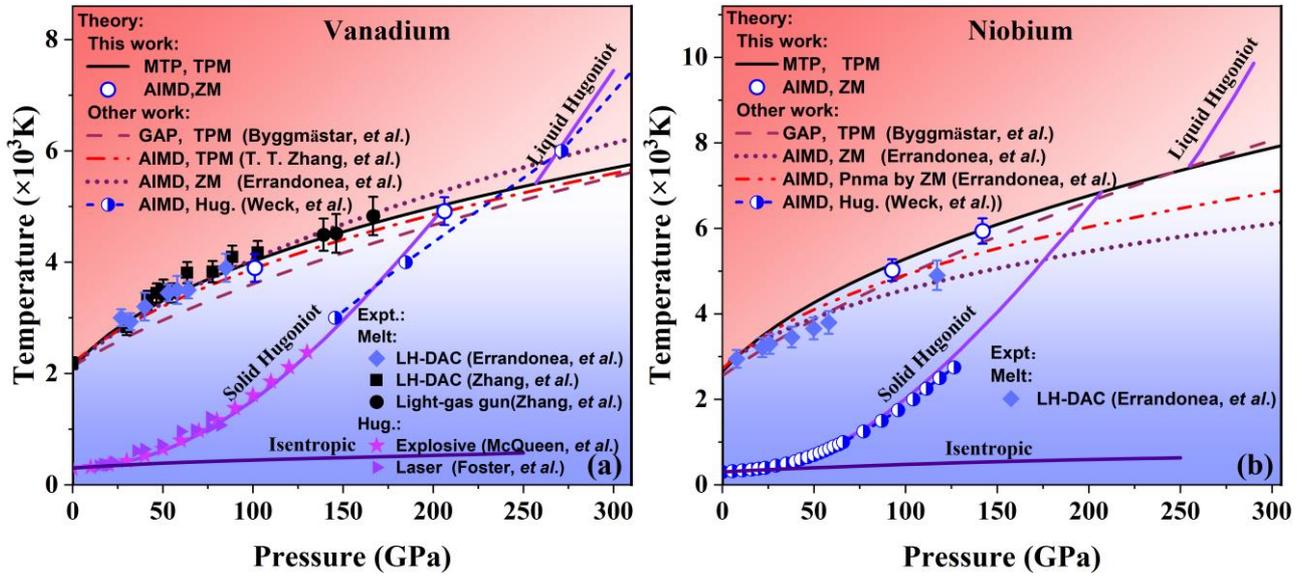

**Fig. 7**. (Color online) Melting and Hugoniot curves of bcc V (a) and Nb (b). The black solid line represents the Simon equation fitted based on MTP results. The open circle is the melting result obtained by AIMD. The different dashed lines symbolize other theoretical melting results of V and Nb [5, 7, 32, 68]. The half-filled pattern represents the Hugoniot values calculated by Weck [69, 70]. The solid symbols represent the experimental melting results [5-7] and Hugoniot values [71, 72].

As shown in Fig. 7(b), our MTP-provided melting results for Nb differ significantly from the experimental and theoretical reports by Errandonea [7]. The melting temperature of Nb at ambient pressure is 2652 K, which agrees with the experimental value of 2750 K [7]. Our initial Clapeyron slopes of BCC Nb are 47.1 K/GPa, which is comparable with Fellinger's classical potential results (50 K/GPa) [21] and differs significantly from other reported results (22–25, 62–65 K/GPa) [7, 73, 74]. By contrast, the laser-heated diamond-anvil cell experiment and AIMD data of Errandonea [7] underestimated the melting temperature of the BCC phase by 10–20% above 70 GPa. We further confirmed this by using the Z method based on AIMD to reproduce the melting values of bcc V and Nb reported in their theoretical calculations, while ensuring consistency in the supercell size and density in our calculations. The results of the final comparisons are presented in Table 2. Our AIMD-Z method predicted a higher melting pressure and temperature at the same density of Nb as the values calculated by Errandonea [7]. Considering that Errandonea has not publicly provided the energy cutoff (ENCUT), we tested the impact of various ENCUTs on the results, all of which were higher than the theoretical values of Errandonea [7] (see the Supplementary Material [65]). This analogous discrepancy in melting curves occurs similarly in the homologous metal Ta. The theoretical melting curve of the metal by AIMD aligns with the experimentally determined shock melting temperature but exceeds the static high-pressure melting curve also measured via LH-DAC [32, 75, 76].

**Table 2**. Melting temperatures and pressures of V and Nb at different densities obtained from the AIMD-Z method. The results obtained by Errandonea *et al*. [5, 7] using the AIMD-Z method at the same density are also included for comparison.

| Z method | $\rho$ (g/cm$^3$) | $T_m$ (K) | | $P_m$ (GPa) | |
|---|---|---|---|---|---|
| | | This study | Errandonea [5, 7] | This study | Errandonea [5, 7] |
| V | 8.13 | 3898 | 4060 | 101 | 101 |
| | 9.62 | 4915 | 5230 | 206 | 207 |
| Nb | 10.88 | 5023 | 4330 | 93 | 79.9 |
| | 11.96 | 5936 | 4820 | 142 | 124 |

The Hugoniot curves of V and Nb were calculated by the Rankine–Hugoniot equation [77] $E$-$E_0$ = -0.5$(P+P_0)(V-V_0)$, where $E$ is the specific internal energy, $P$ is the pressure, and $V = 1/\rho$ is the specific volume of shocked BCC V and Nb. The initial densities were at 0 GPa and 300 K. The results are shown in Fig. 7. The solid and liquid Hugoniot curves of V predicted by the MTP are supported by the current theoretical [70] and experimental results [71, 72]. The Hugoniot solid of Nb matches Weck's AIMD results [69]. Our results also offer a valuable reference for modeling the equation of state in hydrodynamics.

### 3.4. Anomalous Elastic Properties at High Temperature and Pressure

After determining the phase diagrams of V and Nb, this study subsequently conducts a systematic analysis of the temperature-dependent elastic behavior of these metals under high-pressure conditions. Two critical pressures were selected (200 GPa for V and 70 GPa for Nb) to elucidate the distinct effects of pure electron temperature and atomic thermal displacement on the HIH behavior of V and Nb. Compression-induced softening reaches its maximum at these pressures. At these pressures, AIMD simulations are conducted to simulate the elastic properties under two conditions: incorporating electronic temperature effects and assuming electrons remain in their ground state at 300 K reference conditions. For the purpose of comparative analysis, the elastic properties were also calculated using static methods that consider only electronic temperature contributions. These methods were selected in accordance with established protocols from previous studies [12, 13]. The results are presented in Fig. 8.

The results of both sets of AIMD simulations (incorporating and neglecting electronic temperature, respectively) demonstrated consistent shear modulus behavior for both single-crystal and polycrystalline V and Nb, with deviations of approximately 10 GPa observed at elevated temperatures. Static calculations predicted continuous shear instability in single-crystal V up to 1500 K, whereas AIMD simulations indicated suppression of this instability above 300 K. This demonstrates that atomic thermal displacements effectively inhibit shear instability in these systems. Notably, the temperature dependence of the shear elastic constant $C'=(C_{11}-C_{12})/2$ in static calculations diverges from that predicted by AIMD at temperatures above 1500 K. This prediction is incongruent with both comprehensive AIMD results and experimentally observed $C'$ evolution at ambient pressure (Figs. 8(a), 8(d), and Fig. 4). This discrepancy confirms that atomic thermal displacements dominate the HIH of the shear moduli in V and Nb under realistic conditions, in sharp contrast to static predictions based on ideal lattice models. The early theoretical study on V [78] and Nb [79] indicated that atomic thermal displacements and electronic temperature suppress electron–phonon coupling-induced softening of phonon spectra, leading to phonon hardening. Although elastic properties were not directly discussed, according to the long-wavelength limit ($\lim_{|k|\to 0} d\omega/dk = C$, where $C$ represents the single-crystal sound velocity), the transverse acoustic branch hardening near the $\Gamma$ point reflects enhanced $C_{44}$ ($C_{s1} = C_{s2} = \sqrt{C_{44}/\rho}$, where $C_{S1}$ and $C_{S2}$ represent the single-crystal shear sound velocity).

AIMD further reveals a transition from HIH to HIS in the shear moduli ($G$) of polycrystalline V (Nb) at elevated temperatures. As shown in Fig. 8(c) and (f), since the shear sound velocity is directly related to the shear modulus ($C_S^2 = G/\rho$), the shear sound velocity thus exhibits the same behavior. This transition is attributed to the predominance of thermal expansion-induced softening, which suppresses HIH behavior. A comparison of the MTP predictions with the AIMD results reveals minor numerical discrepancies while maintaining substantial agreement in the description of the mechanical behavior of V and Nb. These discrepancies are primarily attributable to potential function-fitting errors that occur during the training process. In such cases, highly diverse atomic spatial environments can result in larger stress calculation uncertainties.

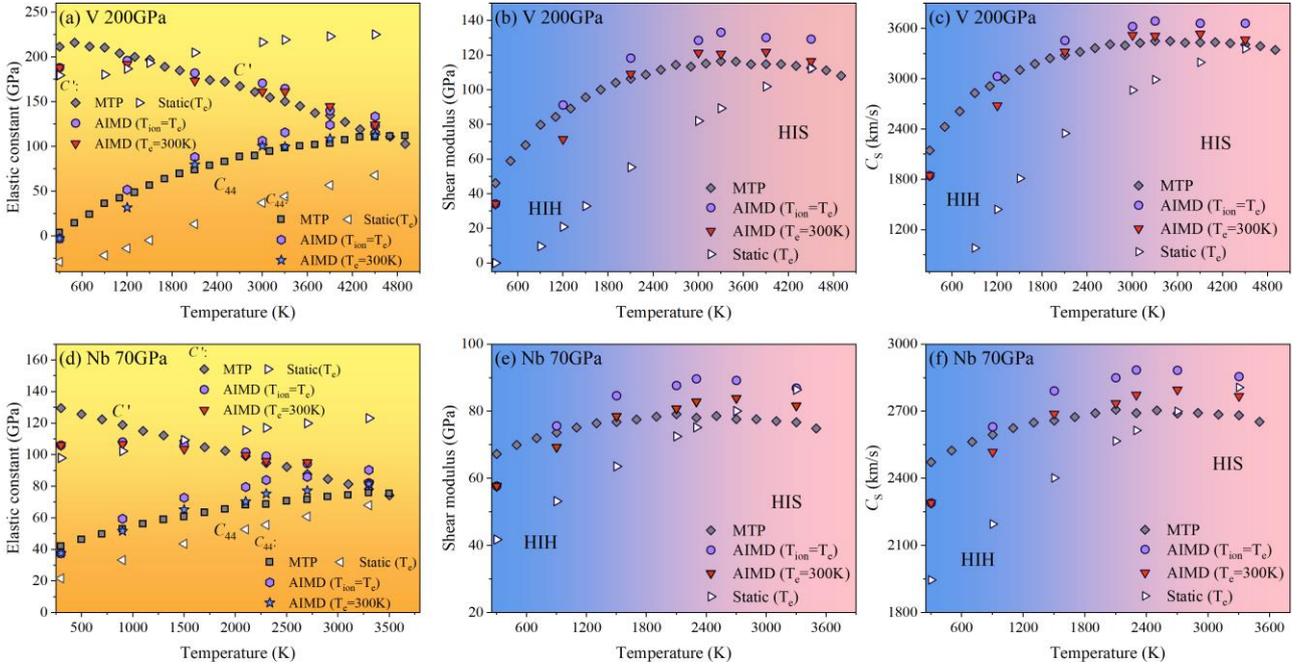

**Fig. 8**. (Color online) Calculated variation of the elastic constants and sound velocities of BCC V and Nb as a function of temperature at 200 and 70 GPa. (a) $C_{44}$ and $C'$ of V; (b) $G$ of V; (c) $C_S$ of V; (d) $C_{44}$ and $C'$ of Nb; (e) $G$ of Nb; (f) $C_S$ of Nb. Solid circles and hexagons represent AIMD results with ion–electron temperature coupling. Pentagrams and downward triangles represent AIMD results considering only ion temperature variation. Hollow triangles represent static calculation results considering only electron temperature. Solid squares and diamonds represent MTP results.

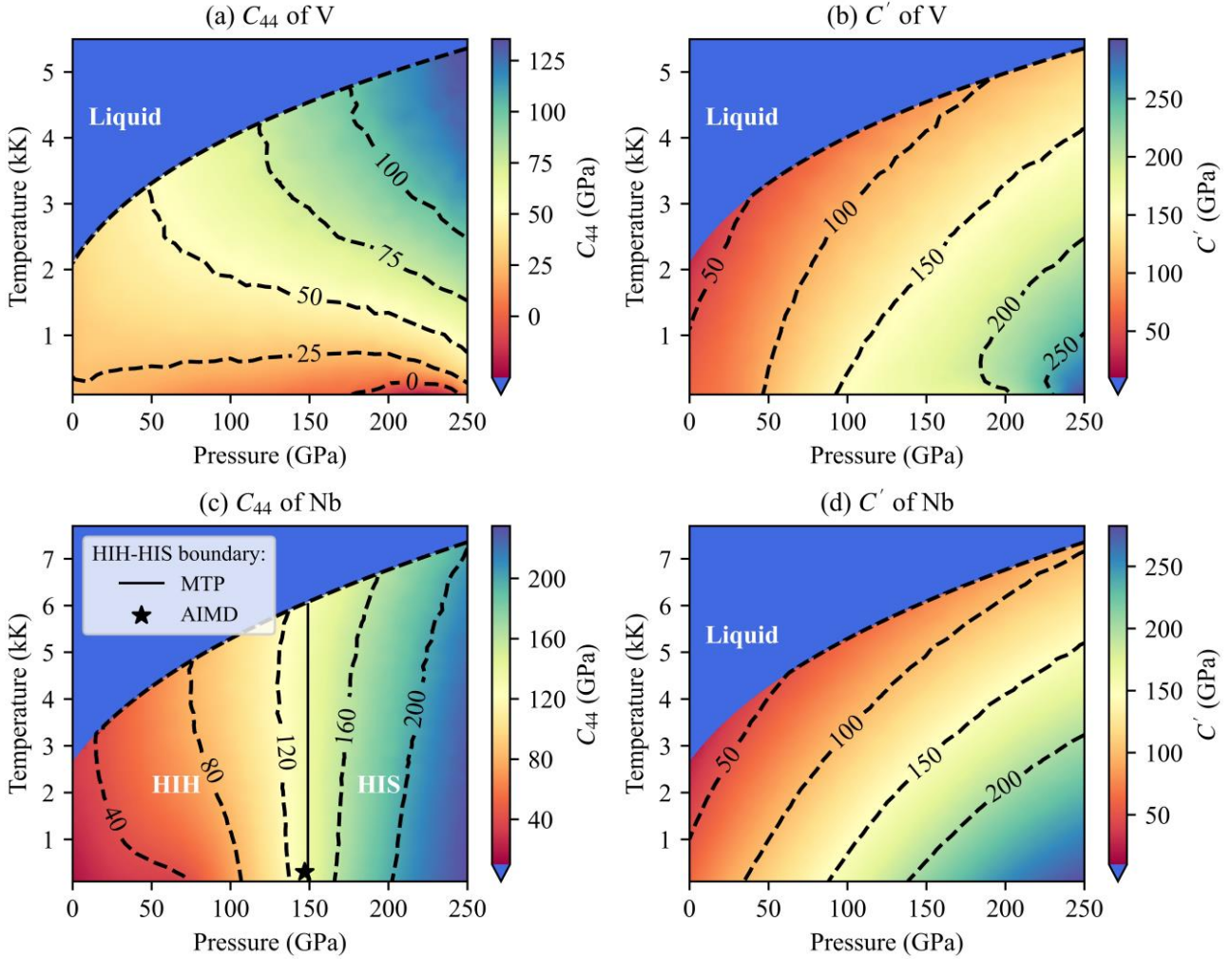

**Fig. 9**. (Color online) Calculated variation of single-crystalline elastic constants of BCC V and Nb as a function of hydrostatic pressure and temperature. (a) $C_{44}$ of V; (b) $C'$ of V; (c) $C_{44}$ of Nb; (d) $C'$ of Nb. The color map indicates the value of the elastic constants.

We performed molecular dynamics simulations using the MTP to investigate the elastic properties of V and Nb as functions of temperature and pressure. The results of the shear elastic constants $C_{44}$ and $C'$ for single-crystalline V and Nb are shown in Fig. 9. Our calculations confirm the previously predicted anomalous softening–hardening effect in $C_{44}$ for single-crystalline V and Nb. At low temperatures, the $C'$ of V exhibited weak CISHIH behavior within the pressure range of 200–250 GPa. Fig. 9(a) shows that the shear instability pressure of BCC V was approximately 200 GPa at 300 K, significantly lower than the previously reported theoretical range of 80–125 GPa at 0 K. This difference can be attributed to the shielding effect of atomic thermal displacements on the anomalous electronic behavior of V. Hence, the phase transition pressure from the BCC to the RH phase driven by shear softening could have been delayed, which explains the experimental findings of Akahama [54] that V remained stable in the BCC phase at 300 K up to 200 GPa. Notably, other theoretical studies [8, 10-13] were based on idealized lattice structures. Furthermore, both $C_{44}$ samples maintained the HIH effect near the melting temperature. Fig. 9(c) shows that the $C_{44}$ of Nb underwent a transition from HIH to HIS when it exceeded 150 GPa, which matches the critical value predicted by AIMD. This may be associated with the disappearance of electronic topological transition behavior in Nb. In contrast to $C_{44}$, the shear elastic constants $C'$ of V and Nb exhibited pronounced temperature dependence, softening rapidly with increasing temperature.

The results are presented in Fig. 10(a) and (d), which show that the bulk moduli of V and Nb did not exhibit similar anomalies; the bulk modulus of V was less sensitive to temperature changes under high pressure. The $G$ and Young's modulus ($E$) (Fig. 10(b) and (c)) of V exhibited significantly anomalous CISHIH behavior. Similar to $C_{44}$, the $G$ and $E$ values of V decreased with pressure at lower temperatures, reaching a minimum of 220 GPa. By contrast, no significant CIS anomalies were observed in the $G$ (Fig. 10(e)) and $E$ (Fig. 10(f)) of Nb, which is inconsistent with the results obtained at zero temperature [13]. This may be attributed to the fact that the hardening effect due to temperature was significantly stronger than that of the CIS. As the temperature increased, $G$ and $E$ hardened. However, once a certain temperature was reached, the hardening behavior caused by the temperature-dependent electronic structure [78, 79] was overshadowed by the softening effect resulting from the lattice thermal expansion. This led to a shift from HIH to HIS. The transition temperatures of V were 1200 K at 0 GPa and 3300 K at 250 GPa, with the overall transition temperature approximately 60% of the melting temperature. The HIH region of Nb was relatively narrow, disappearing entirely at 130 GPa, with a maximum transition temperature of approximately 2500 K at 80 GPa. As can be seen from the theoretical definition of sound velocity ($C_S^2 = G/\rho$, $C_b^2 = B/\rho$) [80-82], the sound velocity behavior of V and Nb is consistent with their polycrystalline modulus behavior. As shown in Figs. S3 and S4 of the Supplementary Material [65] (see also [83-86] therein), the shear sound velocity also exhibits the same HIH–HIS transition.

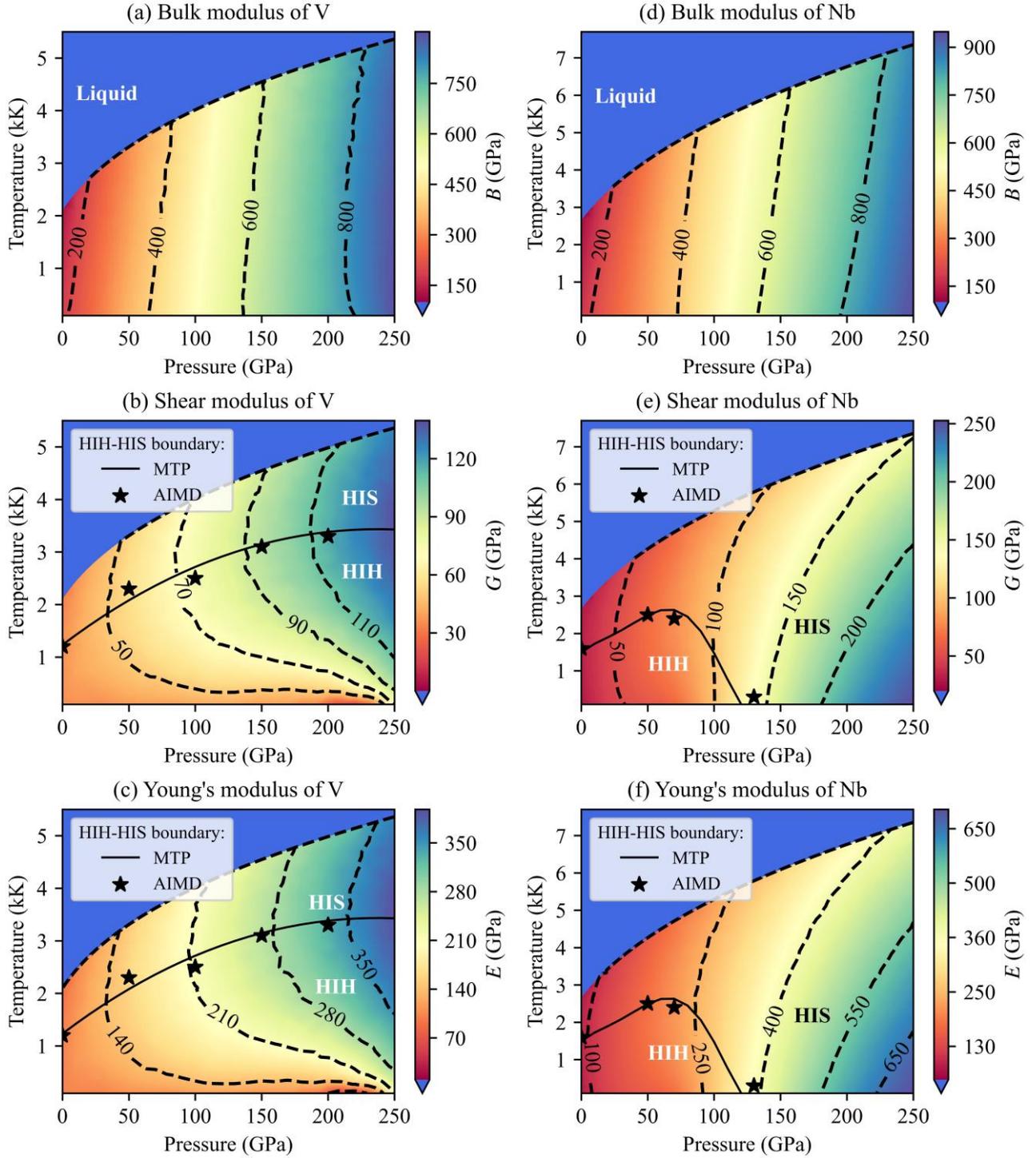

**Fig. 10**. (Color online) Bulk modulus ($B$), Young's modulus ($E$), and shear modulus ($G$) of (a–c) V and (d–f) Nb as a function of temperature and pressure. The black solid line and pentagram represent the transition temperature from HIH to HIS predicted by MTP and AIMD under different pressures. The color map indicates the value of the modulus.

## 4. Conclusions

In this study, we adopted the on-the-fly generation strategy of *VASP* to construct a DFT dataset of V and Nb and train the MTP potentials. The resulting MTP demonstrated high accuracy and robustness comparable with that of DFT. Based on AIMD simulations, we propose that the high-temperature Pnma phase of Nb reported in experiments is highly susceptible to mechanical instability at elevated temperatures, causing reversion to the BCC phase. Theoretically, the stable existence of the Pnma phase may stem from specific constraints along principal stress directions. In addition, the MTP was used to predict the melting and Hugoniot curves of V and Nb under high pressure. The high-pressure melting results of Nb predicted by the MTP and AIMD methods were significantly higher than those of previous experimental and DFT reports.

AIMD with ion–electron decoupling reveals that atomic thermal displacements, not electron temperature, dominate the thermal hardening in V and Nb at high temperatures—contrary to prior theories. We also predicted the pressure and temperature variations in the elastic constants and sound velocities for single crystals and polycrystals of V and Nb. The $C_{44}$ of single-crystal V and Nb exhibits features characteristic of CISHIH. The Young's modulus, shear modulus, and shear sound velocity of polycrystalline V and Nb generally exhibit anomalous mechanical softening behaviors. As the temperature increases, the mechanical properties of V and Nb transition from HIH to HIS. This suggests competition between the anomalous electronic structure behavior and the lattice softening effects. Our research enhances the understanding of the intricate high-pressure characteristics of V and Nb, offering a theoretical foundation for subsequent precise experimental designs of these materials.

## Competing Interests

The authors declare that they have no competing financial interests or personal relationships that may have influenced the work reported in this study.

## Acknowledgments

This work was supported by the National Key R&D Program of China under Grant No. 2021YFB3802300, the NSAF under Grant No. U1730248, and the National Natural Science

Foundation of China under Grant Nos. 12274381 and 12404287. The simulation was performed using resources provided by the Center for Comput. Mater. Sci. software (Tohoku University, Tokyo, Japan).

## Author Contributions

**Hao Wang**: Software, Formal analysis, Visualization, Writing of the original draft. **Dan Wang**: Formal analysis and validation. **Long Hao**: Formal analysis. **Jun Li**: Writing, review, formal analysis, and funding acquisition. **Hua Y. Geng**: Writing, review and editing, and funding acquisition.

# Supplementary material

# High-pressure melting and elastic behavior of vanadium and niobium

# based on *ab initio* and machine learning molecular dynamics

Hao Wang[1], Dan Wang[1, 3], Long Hao[1], Jun Li[1], Hua Y. Geng[1, 2*]

[1] *National Key Laboratory of Shock Wave and Detonation Physics, Institute of Fluid Physics, CAEP, Mianyang 621900, People's Republic of China*

[2] *HEDPS, Center for Applied Physics and Technology, and College of Engineering, Peking University, Beijing 100871, People's Republic of China*

[3] *Institute of Atomic and Molecular Physics, College of Physics, Sichuan University, Chengdu 610065, People's Republic of China*

## A. Test of melting results of niobium under AIMD-Z method

We investigated the effect of different ENCUT on the melting results of Nb. As shown in Table S1, our calculated results are all higher than those calculated by Errandonea[1, 2].

**Table S1**. The melting results of Nb obtained by different ENCUT are shown in the table. The calculation results of Errandonea[1, 2] are also listed in the table.

| $\rho$ = 11.96 (g/cm$^3$) | 250 eV | 300 eV | 400 eV | 500 eV | Errandonea[1, 2] |
|---|---|---|---|---|---|
| $T_m$ (K) | 5876 | 5893 | 5642 | 5936 | **4820** |
| $P_m$ (GPa) | 125 | 137 | 141 | 142 | **124** |

## B. The melting process of Pnma-niobium under AIMD-Z method

The structural transformation of Pnma-Nb during its melting process has been presented using the Z method at a particular density. The supercell of BCC (Pnma) Nb contains 432 (448) atoms. Figures S1 and S2 present the calculated XRD patterns and radial distribution function of the Pnma and BCC phases. During the NVE simulation, XRD indicates a significant phase transition in the initial Pnma phase, whereas no change is observed in the initial BCC phase. Furthermore, Fig. S2 reveals that the radial distribution functions of the Pnma and BCC phases are highly similar at elevated temperatures. Consequently, it is postulated that when utilizing the Z-method to estimate the melting temperature of the initial Pnma phase, the phase has undergone a transformation into other phases prior to melting, potentially the BCC phase.

* *Corresponding author. E-mail: s102genghy@caep.cn*

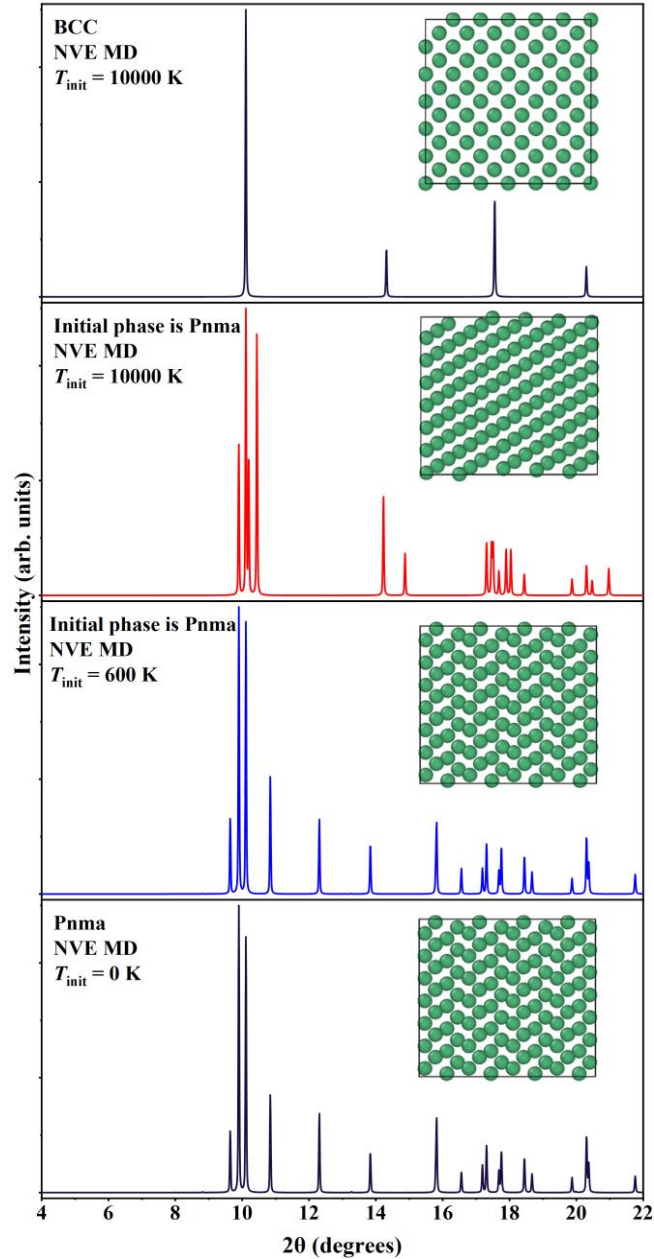

**Fig. S1**. (Color online) The calculated XRD patterns of Pnma-Nb and BCC-Nb at different temperatures based on NVE AIMD.

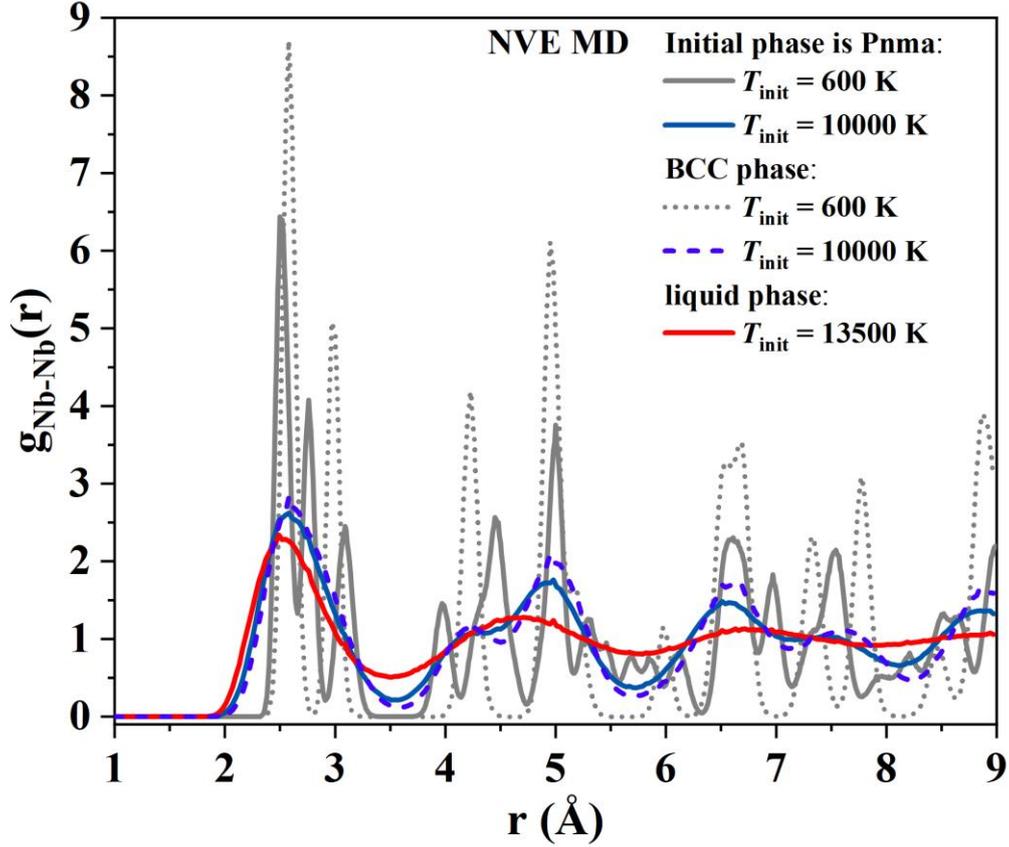

**Fig. S2**. (Color online) The radial distribution functions of the BCC phase, Pnma phase, and liquid phase of Nb are provided based on NVE AIMD.

### C. Anomalous sound velocity behavior

In experiments, the direct measurement of the elastic modulus of a material is challenging, particularly under extreme conditions. The elastic modulus is closely linked to the sound velocity of materials. Therefore, dynamic high-pressure experiments often rely on the sound velocity of materials to study changes in mechanical properties and melting [3]. Fig. S3 shows the sound velocity behavior of V and Nb through the polycrystalline modulus and material density. The bulk sound velocity ($C_b^2=B/\rho$) behavior matches the bulk modulus. The shear sound velocities ($C_s^2=G/\rho$) of V and Nb exhibited distinct CISHIH characteristics, with their transition pressure and temperature from HIH to HIS matching that of the shear modulus. Except in the low-temperature range under high pressure, the longitudinal sound velocity ($C_L^2 = C_b^2+4C_s^2/3$) of V did not exhibit significant CISHIH characteristics. The $C_L$ of Nb was minimally influenced by temperature within 125 GPa but showed pronounced HIS behavior at higher pressures.

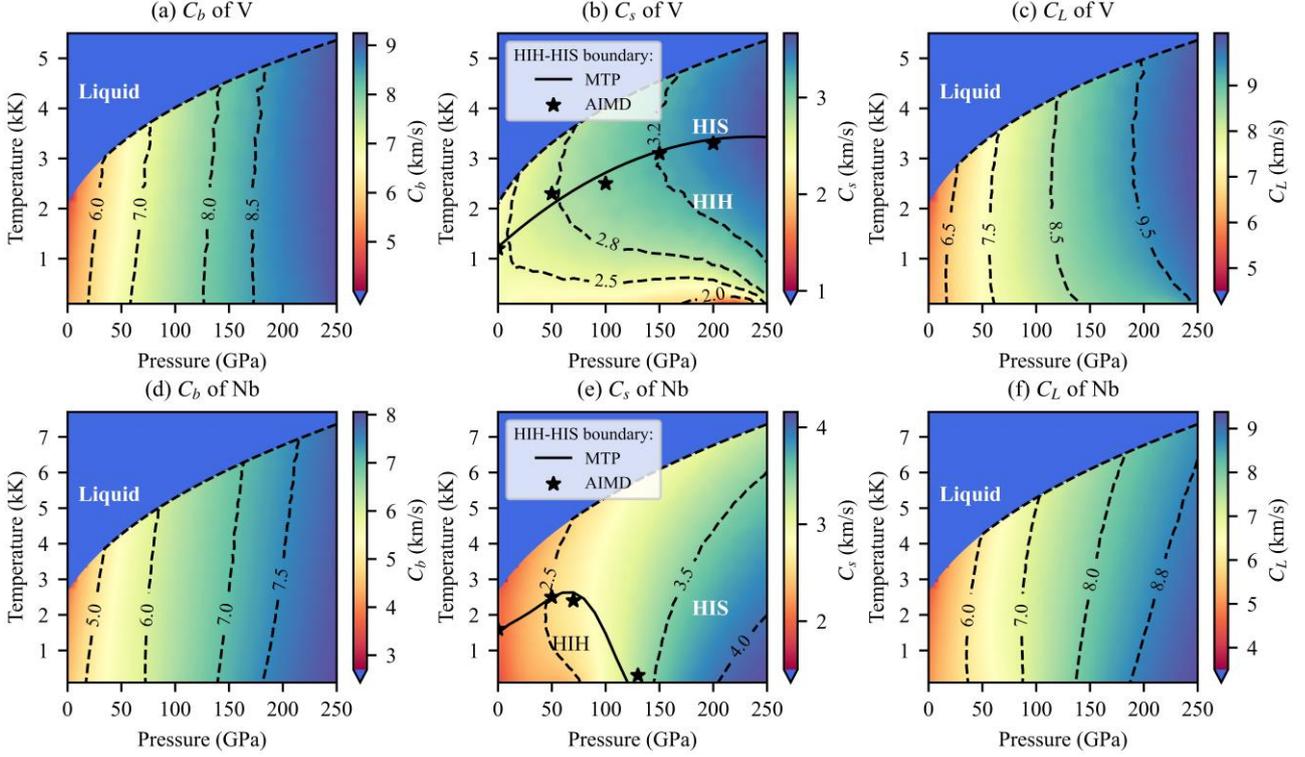

**Fig. S3**. (Color online) Bulk sound velocity ($C_b$), shear sound velocity ($C_s$), and longitudinal sound velocity ($C_L$) of V (a–c) and Nb (d–f) as a function of temperature and pressure. The black solid line and pentagram represent the transition temperature from HIH to HIS predicted by MTP and AIMD under different pressures. The color map indicates the value of sound velocity.

Impact loading is commonly used for high-pressure sound velocity measurements. We present a further comparison between the theoretically predicted shock sound velocity and the existing experimental results [3-6]. Because the $C_b$ of V is weakly affected by temperature, we used the theoretical $C_b$ as a reference and derived the $C_s$ of Dai [3] and Wang [5] from the directly measured $C_L$ ($C_L{}^2 = C_b{}^2 + 4C_s{}^2/3$). Fig. S4(a) demonstrates that the shock sound velocities of V predicted by the MTP align with the experimental values [3-5]. Owing to the coupling between the shock pressure and temperature, the CIS effect of $C_s$ softening was weaker than that of the static high pressure. The sound velocity of polycrystalline Nb (Fig. S4(b)) predicted by the MTP was lower than the experimental results of Li [6]. According to Li's experimental results, the $C_L$ of Nb at 70 GPa is projected to be 8.0 km/s, significantly exceeding the theoretical prediction of MTP (6.7 km/s), and the shock experimental results of V (approximately 7.6 km/s) [5] and Ta approximately 5.3 km/s [7] in the same group-VB. Furthermore, the MTP results did not exhibit the anomalous softening behavior reported by Li within the pressure range of 50–70 GPa.

Moreover, the trend in the shock sound velocity reported in Li's experiment with pressure deviated from the MTP prediction. To better discuss the differences between experiment and theory, we derived the $C_b$ behavior of V and Nb along the Hugoniot using existing experimental data and the shock Hugoniot equation of state (EOS). The relevant formulas are as follows [8, 9]:

$$U_s = C_0 + \lambda U_p$$

$$\gamma V_0 = \gamma_0 V$$

$$P = \rho_0 U_s U_p = \frac{C_0^2}{V_0} \frac{1 - \dfrac{V}{V_0}}{\left[ 1 - \lambda \left( 1 - \dfrac{V}{V_0} \right) \right]^2}$$

$$C_b^2 = V^2 \left[ \left( \frac{\partial P}{\partial V} \right)_H \left[ (V_0 - V) \frac{\gamma}{2V} - 1 \right] + \frac{P\gamma}{2V} \right]$$

where $U_s$ ($U_p$) is shock wave velocity (particle velocity), $C_0$ and $\lambda$ are Hugoniot parameters, $\rho_0$ is initial density, $V = \rho^{-1}$ is specific volume, and $\gamma$ is the Grüneisen coefficient. The $U_s$ ($U_p$), $C_0$, and $\lambda$ data of V and Nb were obtained from the shock experiment of McQueen [10] and Marsh [11]. The $\rho_0$ and $\gamma_0$ of V (Nb) were set to experiment values of 6.11 (8.57) g/cm$^3$ and 1.35 (1.27) [12, 13]. Fig. S4 shows that the $C_b$ calculated by the experimental EOS was consistent with the results of the MTP. In contrast, the trend of $C_b$ of Nb measured by Li deviated from the results derived from the experimental EOS and MTP calculations. Therefore, we may require more precise experimental measurements to reexamine the current MTP results and the experimental results of Li *et al*. [6].

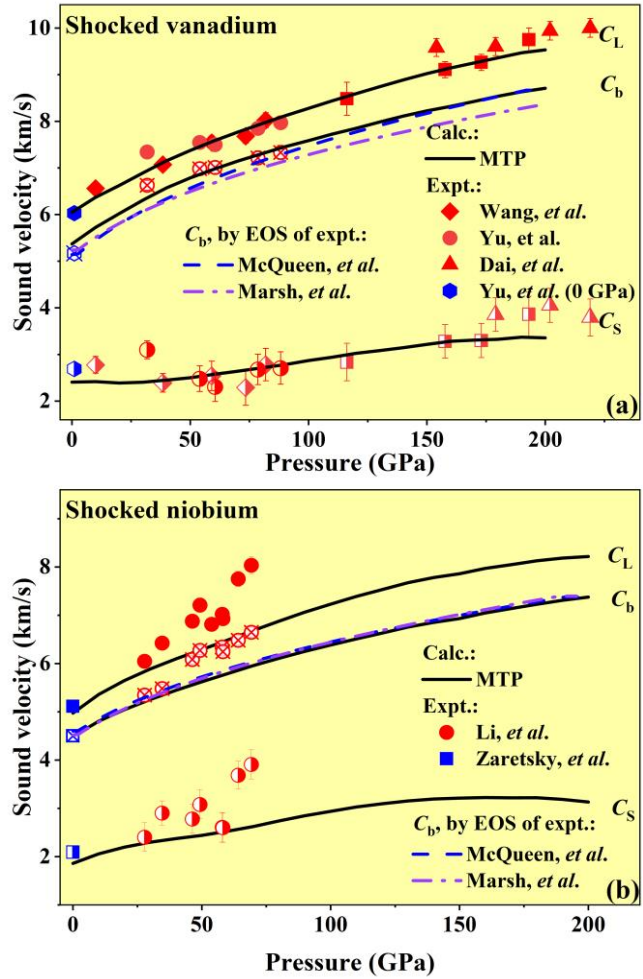

**Fig. S4**. (Color online) Calculated sound velocity of shocked V and Nb along the Hugoniot. The experimental sound velocities of Wang [5], Dai [3], Yu [4], Li [6], and Zaretsky [14] are also given. The solid point and half-filled point represent $C_L$ and $C_S$, respectively. The $C_S$ of Dai [3] and Wang [5] are given through experimental $C_L$ and theoretical $C_b$. The open crossed red points of V and Nb are for $C_b$, as reported by Yu *et al.* [4] and Li *et al* [6], respectively. The dashed (dot-dash) line represents the $C_b$ mainly derived from the experimental $U_s$, $U_p$ of McQueen (Marsh) [10, 11].